# Solving the stationary Liouville equation via a boundary element method

David J. Chappell, Gregor Tanner

*School of Mathematical Sciences, University of Nottingham, University Park, Nottingham NG7 2RD, UK*

## Abstract

Intensity distributions of linear wave fields are, in the high frequency limit, often approximated in terms of flow or transport equations in phase space. Common techniques for solving the flow equations for both time dependent and stationary problems are ray tracing or level set methods. In the context of predicting the vibro-acoustic response of complex engineering structures, reduced ray tracing methods such as Statistical Energy Analysis or variants thereof have found widespread applications. Starting directly from the stationary Liouville equation, we develop a boundary element method for solving the transport equations for complex multi-component structures. The method, which is an improved version of the Dynamical Energy Analysis technique introduced recently by the authors, interpolates between standard statistical energy analysis and full ray tracing, containing both of these methods as limiting cases. We demonstrate that the method can be used to efficiently deal with complex large scale problems giving good approximations of the energy distribution when compared to exact solutions of the underlying wave equation.

## 1. Introduction

Many phenomena in physics and engineering can be accurately described or modelled in terms of linear wave equations. A large range of numerical methods has been developed for solving wave problems, often making heavy use of the linearity of the underlying PDEs leading in general to large, finite matrix equations. Popular tools include finite element or finite volume methods, boundary element methods and various spectral methods. There are, however, basic limitations when approximating the solutions of the wave equations directly: the size of the associated linear system increases with decreasing wavelength and numerical schemes become inefficient when the local wavelengths are orders of magnitude smaller than typical dimensions of the physical system.

In order to overcome this limitation, a range of high-frequency/ short wave length methods have been devised such as the WKB or eikonal approach or semiclassical methods developed in the context of quantum chaos, see [1, 2, 3] for overviews. In particular, phase information is obtained by solving the Hamilton-Jakobi equations for the action fields, whereas amplitude information is calculated using transport equations. The actual methods can be quite intricate demanding an intimate knowledge of the classical phase space dynamics given by the Hamilton equations of motion for the underlying ray or classical dynamics. If one is prepared to omit phase information altogether, the approximations amount to solving the transport or Liouville equation for phase space densities.

In this paper, we will focus on efficient numerical methods for solving the Liouville equation (LE) for stationary problems. We will show how the solutions of the LE are related to both ray tracing (see for example, [4, 5, 6]) and *statistical energy analysis* (SEA) [7, 8, 9], a popular method in the engineering community for solving vibro-acoustic problems in the high-frequency limit.

The Liouville equation is an example of a PDE describing a conservation law [10]. Typical solution methods are based on characteristics or ray tracing. Ray tracing is the method of choice when considering propagation over relatively short times or length intervals. Prime examples include weather forecasting, where data updates lead to regular re-initialisations of the density distributions [11], or room acoustics, where only very few reflections need to be considered [5]. More sophisticated methods related to tracking the time-dynamics of interfaces in phase space such as moment methods and level set methods [12, 13, 14] have been developed only relatively recently, finding applications in acoustics, seismology and computer imaging. For an excellent recent overview, see Runborg [2]. Numerical methods based on solving the time dependent Liouville equation directly on a mesh using finite volume or transfer operator methods, also referred to as Ulam methods, have been considered in [15] and more recently in [16, 17]. Ray based and tracking methods often become inefficient when considering frequency domain wave problems in bounded domains, for example, determining the wave field in a finite size cavity driven by a continuous, monochromatic excitation. Here, multiple reflections of the rays and complicated folding patterns of the associated level-surfaces often lead to an exponential increase in the number of branches that need to be considered.

In the engineering literature, these problems have been circumvented by subdividing the structure into a set of substructures. Assuming ergodicity of the underlying ray dynamics and quasi-equilibrium conditions in each of the subsystems, that is, assuming that the density in each subsystem is approximately constant, greatly simplified SEA equations based only on coupling constants between subsystems can be set up. The disadvantage of SEA is that the underlying assumptions are often hard to verify *a priori* or are only justified when an additional averaging over 'equivalent' subsystems is considered. These shortcomings have been addressed by Langley [18, 19] and more recently in a series of papers by Le Bot [20, 21, 22]. A computational tool based on a Frobenius-Perron operator approach systematically interpolating between SEA and full ray tracing has been introduced first in [23] and further analysed in [24]. Its name, *Dynamical Energy Analysis* (DEA), points at the similarities with SEA but stresses at the same time the importance of dynamical correlations present in the ray-densities. Thus, in DEA, the solution in a subsystem of the structure is typically represented by a vector; this is in contrast to SEA where the solution in each subsystem is represented by a single number, the mean ray density in each subsystem.

The implementation of DEA as presented in [23, 24] corresponds to a spectral boundary integral method; the integral equations are expanded in orthogonal basis sets and the resulting matrix equations are solved for the coefficients of the basis expansion of the solution. It has been shown in [24] that the method works even in situations where a naive application of SEA fails. However, spectral methods can lead to slow convergence for the matrix elements in the DEA matrix, in particular if the boundary is non-smooth. In this paper, we show that using a boundary element method (BEM) for the spatial variable and a basis function expansion in the momentum component leads to large efficiency gains. We will furthermore present a more in depth derivation of the actual boundary integral formulation.

The paper is structured as follows: in Sec. 2, we sketch the derivation of the Liouville transport equation from the underlying wave equation; we will then derive the boundary integral formulation in Sec. 3 and give a description of the numerical implementation of DEA using the BEM/ basis function approach in Sec. 4. Numerical results will then be presented in the Sec. 5.

## 2. From wave dynamics to phase space densities

### 2.1. A time dependent formulation

In the following, we sketch the connection between linear wave equations and the Liouville equation via the Eikonal ansatz. We develop the theory in the time domain first and move to a stationary formulation in the next section. We will restrict the discussion to the Helmholtz equation of acoustics; other wave problems such as electromagnetic waves, linear elasticity or Schrödinger's equation can be treated along the same line with the obvious modifications. For a more in depth discussion of the Eikonal approach, see for example [2]. A nice derivation of the connection between wave transport equations and the Liouville equation based on Wigner transformation techniques has been given in [25].

We start from the wave equation for linear acoustics, that is,

$$\left(\frac{\partial^2}{\partial t^2} - \nabla \cdot (c^2 \nabla)\right) u(r, t) = 0 \tag{1}$$

seeking solutions $u(r, t)$ in some domain $\Omega$ with boundary $\Gamma$ ($u$ may express acoustic pressure in appropriate units). Furthermore, $c = c(r) \geq 0$ is the local wave speed. We assume in what follows that the wave equation is not explicitly time dependent and that the walls act as passive elements, that is, we have Dirichlet or Neumann boundary conditions, $u(s) = 0$ or $\partial u/\partial n(s) = 0$ for $s \in \Gamma$. Solutions are then determined uniquely in terms of initial conditions $u(r, 0) = u_0(r)$ and $\partial_t u(r, 0) = u_1(r)$ at $t = 0$.

We now make the usual Eikonal ansatz

$$u = A e^{i\omega\phi} \tag{2}$$

with real valued functions $A = A(r, t), \phi = \phi(r, t)$ and $\omega$ acting as the large parameter here. Plugging (2) into (1) and collecting terms of the order $\omega^2$, we arrive at the Eikonal equation

$$\left(\frac{\partial \phi}{\partial t}\right) \pm c |\nabla \phi| = 0. \tag{3}$$

Terms of order $\omega$ yield a transport equation

$$2\frac{\partial A}{\partial t}\frac{\partial \phi}{\partial t} + A\frac{\partial^2 \phi}{\partial t^2} - 2c^2 \nabla A \cdot \nabla \phi - c^2 A \Delta \phi - 2c \nabla \cdot (cA \nabla \phi) = 0. \tag{4}$$

Setting

$$\tilde{\rho}(r, t) = A^2(r, t)\frac{\partial \phi}{\partial t}(r, t),$$

we can write Eq. (4) in the form of a conservation law, that is,

$$\frac{\partial \tilde{\rho}}{\partial t} + \nabla \cdot (v\tilde{\rho}) = 0 \tag{5}$$

with "velocity field"

$$v(r,t) = -c^2 \nabla \phi / \frac{\partial \phi}{\partial t} \qquad (6)$$

determined through the phase function $\phi(r,t)$ alone.

Solutions of the Eikonal equation (3) can readily be found using characteristics. Without restricting the discussion we choose the "+" sign in (3); see also comments further down in the text. Setting

$$H(r,p) = c|p| \quad \text{with momentum} \quad p = \nabla \phi \,, \qquad (7)$$

one obtains solutions $\phi(r,t)$ along trajectories in *phase space* $X(t) = (r(t), p(t))$ by solving Hamilton's equations of motion, that is,

$$\begin{aligned}
\dot{r} &= \nabla_p H = c \frac{p}{|p|} \\
\dot{p} &= -\nabla_r H = |p| \nabla c \,.
\end{aligned} \qquad (8)$$

Note that in deriving the Eikonal Eq. (3) and transport Eq. (4) we omitted terms of order $\mathcal{O}(\omega^0)$; neglecting the coupling terms between Eqs. (3), (4) leads to first order differential equations which can formally be solved by using only one of the initial conditions, that is, either $u_0(r)$ or $u_1(r)$. The ODEs (8) are thus solved by using, for example, initial conditions $X(0) = (r_0, p_0 = \nabla \phi_0(r_0))$. A solution for the phase $\phi$ is then obtained by integration, that is,

$$\phi(r,t) = \phi_0(r_0) + \int_0^t p \, \dot{r} \, d\tau$$

where the integral is taken along $X(t) = \varphi^t(X_0)$. Here $\varphi^t(X_0)$ is the phase space flow map obtained from the pair of ODEs (8). In particular we integrate over trajectories starting at $t = 0$ at some point $r_0$ with momentum $p_0 = \nabla \phi_0(r_0)$ and reaching the final point $r$ at time $t$. Here, $\phi_0$ is the initial phase corresponding to the initial conditions $u_0(r_0) = A_0(r_0) \exp(i\omega \phi_0(r_0))$ at $t = 0$. In general, there will be several such initial points from which a trajectory reaches the point $r$ at time $t$. That is, the solution to the eikonal equation (3) at a point $r$ has a multi-sheet structure with a set of $N = N(r,t)$ branches, each giving rise to a phase

$$\phi_j(r,t) = \phi_0(r_j) + \int_0^t p_j \dot{r}_j d\tau, \quad j = 1, ..., N \,. \qquad (9)$$

The integrals are taken over paths $X_j$, $j = 1, ..., N$, reaching $r$ in time $t$ with $X_j(0) = (r_j, \nabla \phi_0(r_j))$. It is this multi-sheet structure which makes solving Eq. (3) difficult in practice [2]. In particular, the number of branches $N(t)$ can increase rapidly with $t$ and for chaotic ray dynamics $N(t)$ actually increases exponentially [1].

We would like to remark that choosing the "−" sign in (3) leads to a sign-change in (8) and gives time-reversed ray trajectories obtained through the time transformation $t \to -t$. The complete solution to the Eikonal equation is thus given by the set of characteristics travelling both in positive and negative time.

We turn now to the solution of the transport Eq. (5); this equation is driven by the solution $\phi(r,t)$ of the Eikonal Eq. (3). It thus suffers in principle from the same multi-valuedness problem as the original equation. Note, however, that the solutions of the ray equations (8) in phase space are determined uniquely for a given initial condition; it is the projection onto $r$ - space which leads to the multi-valuedness of the phase solution. It is thus advantageous to solve the transport equation in phase space. The corresponding equation is the *Liouville equation* (LE)

$$\frac{\partial \hat{\rho}}{\partial t} + \{H, \hat{\rho}\} = 0 \tag{10}$$

where $\{\cdot, \cdot\}$ are the Poison brackets, and $H(r, p)$ is the classical Hamilton function (7). The phase space density

$$\hat{\rho}(r, p, t) = \sum_{j=1}^{N(t)} \tilde{\rho}_j(r, t) \delta(p - \nabla \phi_j(r, t)) \tag{11}$$

indeed solves Eq. (10) if the branches $\phi_j(r, t)$ and $\tilde{\rho}_j(r, t)$ obey the Eikonal and transport equations (3), (5), respectively. (This result can be obtained by inserting (11) into (10), see also [2].) A formal solution of the LE is obtained in terms of a *Frobenius-Perron operator*

$$\hat{\rho}(X, t) = \mathcal{L}^t[\hat{\rho}](X) = \int \delta(X - \varphi^t(Y)) \hat{\rho}(Y, 0) dY = \hat{\rho}(\varphi^{-t}(X), 0), \tag{12}$$

where the last relation is valid for Hamiltonian flows; this relation is used extensively in the ray tracing approach.

The solution of the wave equation (1) is now formally obtained by summing over the solution branches in (9), that is,

$$u(r, t) = \sum_{j=1}^{N} |A_j(r, t)| e^{i\omega \phi_j(r, t) - i\nu_j \frac{\pi}{2}} \tag{13}$$

where $\nu_j$ are Maslov indices [1] due to caustics along the ray $X_j(t)$. As we are not interested in phases in what follows, we will not discuss this issue any further here. An asymptotic treatment of the wave equation using the Eikonal ansatz needs to keep track of the link between phases and amplitudes along the different solution branches. Ray tracing methods are often the only way forward when finding solutions using the Eikonal ansatz. This may be an efficient method for short times and in cases where only a few reflections occur. However, the rapid increase in the number of paths makes ray tracing methods quite cumbersome for large times. The advantages behind the original idea of the Eikonal ansatz, namely replacing a wave equation with highly oscillatory solutions by differential equations for slowly varying quantities such as the amplitude and phase, are lost.

In many applications it is, however, sufficient to find the solution of the transport equations for a given set of initial conditions. The density $\hat{\rho}(X, t)$ contains a great deal of information about the wave solution, and this may often be sufficient for the required purpose. Classical ray tracing, that is ray tracing without tracking phase information, is indeed used extensively in room acoustics.

### 2.2. Stationary formulation

In what follows, we will consider wave problems with time-harmonic forcing at a frequency $\omega$. We can then separate out time by setting

$$u(r, t) = G(r) e^{i\omega t}.$$

Let us consider problems with a forced excitation by a point source. The corresponding inhomogeneous wave equation then has the form

$$\left(\nabla \cdot (c^2 \nabla) + \omega^2\right) G(r) = -c^2 \delta(r - r_0),\tag{14}$$

with the source point at $r = r_0$. The solution $G(r, r_0)$ is the Green function and arbitrary force distributions can be recovered.

For an Eikonal approach, we proceed by splitting the solution into a homogeneous and an inhomogeneous (or direct) solution in the form

$$G(r, r_0) = G_0(r, r_0) + G_h(r, r_0),$$

where $G_0$ is the free Green function and $G_h$ solves the homogeneous Helmholtz Eq.

$$\left(\nabla \cdot (c^2 \nabla) + \omega^2\right) G_h(r) = 0\tag{15}$$

with appropriate boundary conditions. (For Dirichlet BC, that is $G(s) = 0$, we obtain $G_h(s) = -G_0(s)$ for $s \in \Gamma$.) One finds for two-dimensional problems with $c = const$,

$$G_0(r, r_0) = \frac{i}{4} H_0^1(\omega|r - r_0|/c),\tag{16}$$

where $H_0^1$ is the 0th order Hankel function of the first kind. Proceeding with an Eikonal ansatz, we set $G_h = A \exp(i\omega\psi)$; we have then

$$\phi(r, t) = \psi(r) + t \quad \text{and} \quad \frac{\partial A}{\partial t} = 0.\tag{17}$$

The Eikonal and transport Eqs. (3), (5) reduce to

$$c|\nabla \psi| = 1\tag{18}$$

and

$$\nabla \cdot (v\bar{\rho}) = 0 \quad \text{with } v(r) = -c^2 \nabla \psi.\tag{19}$$

We write the boundary conditions in the form

$$G_0(s) = A_0(s) e^{i\omega\psi_0(s)} \quad \text{for } s \in \Gamma.$$

In order to obtain an approximate solution of (18) at a point $r \in \Omega$ one needs to find those rays starting from a boundary point $s \in \Gamma$ with initial conditions $X_0 = (s, p = \nabla\psi_0(s))$ which pass through the point $r$ following the equations of motion (8). The phase function $\psi$ is then obtained by integrating along this trajectory with an initial condition $\psi(s) = \psi_0(s)$; the phase function $\psi$ again has a multi-sheet structure and there are in general infinitely many sheets intersecting with the plane $r = const$ in phase space. (A phase space description of the sheets can be given in terms of Lagrangian manifolds, see for example [26].) The various branches are calculated according to

$$\psi_j(r) = \psi_0(s_j) + \int_{s_j}^r p \, dr$$

integrating along a trajectory with initial condition $X_j = (s_j, p_j = \nabla \psi_0(s_j))$. The transport equation (19) is solved with initial conditions

$$j_0(s) = -c^2 A^2(s) \frac{\partial \psi(s)}{\partial n} \tag{20}$$

where $j_0(s)$ is the incoming flux normal to the boundary. The total solution is then approximatively obtained as

$$G(r) = G_0(r) + \sum_{j=1} |A_j(r)| e^{i\omega \psi_j(r) - i\mu_j \pi/2} \tag{21}$$

where $\mu_j$ are again Maslov phases and $|A_j|$ is the amplitude transported along the trajectory from $X_j$ to $r$; for alternative derivations of Eqs. (13), (21), see [1].

In applications, one is often interested in wave intensities (such as energy densities); the contributions arising from reflections from the boundary have the form

$$I(r) = |G_h(r)|^2 = \sum_{j,j'} |A_j A_{j'}| e^{i\omega(\psi_j - \psi_{j'})} = \sum_j \tilde{\rho}_j(r) + \sum_{j \neq j'} |A_j A_{j'}| e^{i\omega(\psi_j - \psi_{j'})}. \tag{22}$$

Neglecting the oscillatory terms, which is justified after averaging over a small $\omega$ range or over an ensemble of similar systems, one arrives at the ray tracing approximation

$$I(r) \approx \rho(r) = \sum_j \tilde{\rho}_j(r) = \int \rho(r,p) dp \tag{23}$$

where $\rho(r,p)$ is a solution of the stationary Liouville equation with

$$\{H, \rho\} = 0 \tag{24}$$

and boundary conditions given by (20) together with Snell's law of reflection. Note that the total intensity $|G(r)|^2$ for $r \in \Omega$ may be estimated simply by adding $|G_0(r)|^2$ to $\rho(r)$.

It is clear from the description above that a ray tracing approach is even more cumbersome in the stationary case than in the time-dependent case. The stationary solution is constructed from infinitely many branches which originate from reflections at boundaries. If we neglect phase information, however, we can reconstruct the phase space densities using the Frobenius-Perron operator, Eq. (12) [23, 24]. In the next section we will introduce an efficient algorithm for determining solutions of the *stationary* Liouville equation even in the presence of multiple reflections, thus overcoming the limitations of the ray tracing approaches sketched above.

## 3. Boundary integral formulation of the Liouville equation

In order to solve the stationary Liouville equation (24), the ray tracing integral formula, Eq. (12), is rewritten in a boundary integral form using a boundary mapping technique. The boundary mapping procedure involves first mapping the ray density emanating continuously from the source onto the boundary $\Gamma$ and then constructing a Frobenius-Perron operator for the boundary map.

We now give an explicit derivation of the initial density on the boundary arising from Eq. (14) for Dirichlet boundary conditions. The ray density emanating from a source point $r_0 \in \Omega$ is given by the square of the amplitude of the free space Green function. For $\Omega \subset \mathbb{R}^2$, it is given by (16) and we have the following high frequency asymptotic form

$$|G_0(r, r_0)|^2 \sim \frac{c}{8\pi\omega|r - r_0|}. \tag{25}$$

Lifting the ray density to the full phase space, Eq. (25) corresponds to rays moving out from the source in all directions according to

$$\rho_0(r, p; r_0) = \frac{c\delta(p - p_0)}{8\pi\omega|r - r_0|}, \tag{26}$$

where $p_0 = |p|(r - r_0)/|r - r_0|$. Note that from Eq. (18), the ray dynamics take place on the 'energy' manifold $c|p| = 1$. To obtain the source density $\rho_0^\Gamma$ on the boundary $\Gamma$, we set

$$\rho_0(r, p; r_0) = \rho_0^\Gamma(s, p_s)\delta(c|p| - 1), \tag{27}$$

where $s$ parameterizes $\Gamma$ and $p_s \in B_{|p|}^{d-1}$ denotes the momentum component tangential to $\Gamma$ at $s$ for fixed $H(X) = c|p| = 1$; here $B_{|p|}^{d-1}$ is an open ball in $\mathbb{R}^{d-1}$ of radius $|p|$ and centre $s$. Using local coordinates on the boundary, that is, $p = (p_s, p^\perp)$, then $\delta(p - p_0) = \delta(p_s - p_{s_0})\delta(p^\perp - p_0^\perp)$ and $|p|\delta(p^\perp - p_0^\perp) = cp^\perp\delta(c|p| - 1)$. Combining these identities with (26) and (27) yields

$$\rho_0^\Gamma(s, p_s; r_0) = \frac{c^2\cos(\theta(r(s), r_0))\delta(p_s - p_{s_0})}{8\pi\omega|r(s) - r_0|}, \tag{28}$$

where $\theta(r(s), r_0)$ is the angle between the normal to $\Gamma$ at $r(s)$ and the ray from $r_0$ to $r(s)$. The resulting boundary layer density $\rho_0^\Gamma$ is equivalent to a source density on the boundary producing the same ray field in the interior as the original source field after one reflection.

The operator equivalent to the Frobenius-Perron operator (12), but now acting on boundary densities $\rho^\Gamma$, is described in terms of a boundary integral operator $\mathcal{B}$,

$$\mathcal{B}\rho^\Gamma(X^s) = \int_{\partial\mathbb{P}} w(Y^s, \omega)\delta(X^s - \gamma(Y^s))\rho^\Gamma(Y^s)dY^s \tag{29}$$

where $X^s = (s, p_s)$, $Y^s = (s', p_s')$ and $\gamma$ is the invertible boundary map. Also, $\partial\mathbb{P} = \Gamma \times B_{|p|}^{d-1}$ is the phase space on the boundary at fixed "energy" $H(X) = c|p|$. The weight $w$ contains factors due to reflection/ transmission at interfaces within $\Omega$ and a damping term of the form $\exp(-\mu L(s, s'))$, where $L(s, s')$ is the length of the trajectory between $s$ and $s'$. Note that, depending on the modelling assumptions, the factors in $w$ may depend on $\omega$. A similar weight term also needs to be applied in Eq. (28) since rays emanating from the source will also be subject to dissipation and possible reflection/ transmission at interfaces. We assume uniform damping in the interior, that is, the damping coefficient $\mu$ does not depend on $r$. The damping term does then not alter the underlying ray dynamics [27]. Note that convexity is assumed to ensure $\gamma$ is well defined; non-convex regions can be handled by introducing a cut-off function in the shadow zone as in Ref. [21] or by subdividing the regions further.

The stationary density on the boundary induced by the initial boundary distribution $\rho_0^\Gamma(X^s, \omega)$ can then be obtained using

$$\rho^\Gamma(\omega) = \sum_{n=0}^{\infty} \mathcal{B}^n(\omega)\rho_0^\Gamma(\omega) = (I - \mathcal{B}(\omega))^{-1}\rho_0^\Gamma(\omega), \tag{30}$$

where $\mathcal{B}^n$ contains trajectories undergoing $n$ reflections at the boundary.

The resulting density distribution on the boundary $\rho^\Gamma(X^s, \omega)$ is then mapped back into the interior region. Using Eq. (23), one can relate $\rho^\Gamma$ to a wave energy density after projecting onto $r$ space, that is for $\Omega \subset \mathbb{R}^2$ ,

$$
\begin{aligned}
\rho(r, \omega) &= \int \rho(r, p) dp, \\
&= \int \int \rho^\Gamma(X^s, \omega) \delta(c|p| - 1)|p|d|p|d\Theta,
\end{aligned}
\tag{31}
$$

where $(|p|, \Theta)$ is a polar coordinate system for $p$. One finally obtains

$$
\rho(r, \omega) = \frac{1}{c^2} \int \rho^\Gamma(X^s, \omega) d\Theta.
\tag{32}
$$

Now performing a change of variables from $\Theta$ to $s$ yields

$$
\rho(r, \omega) = \frac{1}{c^2} \int \rho^\Gamma(X^s, \omega) \frac{\cos(\theta(r_s, r))}{|r - r_s|} ds,
\tag{33}
$$

where $r_s \in \Gamma$ is the point on $\Gamma$ in Cartesian coordinates corresponding to $s$. For systems with damping we must again apply a factor of the form $\exp(-\mu L(s, s'))$ to the final density evaluation (33). The total intensity $|G(r, r_0; \omega)|^2$ may be estimated simply by adding $|G_0(r, r_0; \omega)|^2$ to the value of $\rho(r, \omega)$ given by (33).

We remark in passing that our approach is akin to the infinitesimal generator technique presented in [17] which aims at estimating the long term behaviour of the dynamics by numerically calculating the spectral properties of the generator of the Frobenius-Perron operator and extrapolating to long times.

## 4. Numerical implementation: basis approximations and boundary element techniques

The long term dynamics is contained in the operator $(I - \mathcal{B})^{-1}$ and standard properties of Frobenius-Perron operators ensure that the sum over $n$ in Eq. (30) converges for dissipative or open systems. In order to evaluate $(I - \mathcal{B})^{-1}$, a finite dimensional approximation of the operator $\mathcal{B}$ must be constructed. In Refs. [23, 24], basis expansions have been applied in both position and momentum coordinates, which is straightforward to implement for $\Omega \subset \mathbb{R}^2$ using univariate expansions in each argument. In particular, in Ref. [23], a Fourier basis was employed globally on each subsystem in both position and momentum. It is well known that the convergence rate of a Fourier basis expansion depends on the smoothness of the approximant and also requires periodic boundary conditions [28]. The periodicity requirement will be fulfilled due to periodicity of the boundary curve and that the ray density corresponding to $p_s = \pm|p|$ is zero giving periodicity in the momentum coordinate. However, the smoothness criterion often fails if, for example, corners are present in $\Gamma$; the initial density due to a point source is not then differentiable at the corners.

In Ref. [24] it was suggested to split up the spatial approximation at corners in $\Gamma$ and apply a basis approximation with no requirement for periodic boundary conditions in order to attain its optimal convergence properties. An approximation using a Chebyshev basis was implemented since Chebyshev polynomials may be computed simply in terms of trigonometric functions and have similar convergence properties to Fourier basis approximations without the periodicity requirement. Such an approach also has a natural and optimal (in terms of polynomial degrees that can be integrated exactly) quadrature choice for the associated integrals, namely, Gauss-Chebyshev quadrature. The improvement in computational efficiency brought about by this approach opened up the method to larger scale multi-component problems.

In this work we go further along this route by subdividing the boundary using a mesh. This introduces greater flexibility into the numerical method making it possible to both increase the polynomial order of the basis functions and to refine the mesh for the spatial coordinates. The advantage of refining the mesh is that additional degrees of freedom are introduced to the model whilst maintaining relatively low quadrature costs due to the lower order of the spatial basis approximation. As such it is preferable to choose a basis which is orthogonal in the standard $L^2$ inner product. (From previous experience using a Chebyshev basis, where the associated orthogonal inner product is weighted, we found that low order approximations were very poor). Due to the mesh discretization procedure adopted here, the boundary conditions are not periodic and using a Fourier basis is not advisable; hence an alternative $L^2-$orthogonal basis choice is proposed, namely, Legendre polynomials.

The overall form of the proposed numerical scheme in 2D is therefore based on a Legendre polynomial approximation in both position and momentum augmented with a boundary mesh through which the spatial approximation may be localised. For an extension of the method to 3D cavities, see [29]. Recall that $p_s \in (-|p|, |p|)$ and denote by $\tilde{p}_s$ a re-scaling of $p_s$ to $(-1, 1)$. Let us also denote

$$\tilde{P}_m(p_s) = \sqrt{1/|p|} P_m(\tilde{p}_s).$$

An important special case of the scheme described above is when we fix the spatial basis approximation at zero (constant) order and only refine the mesh, which essentially results in a scaled piecewise constant boundary element approximation. This type of approximation is also often referred to as Ulam's method [15, 16], although this would involve performing such an approximation in full phase space, rather than just in its spatial component. The implementation of separate meshes and basis approximations in both position and momentum would, however, complicate the procedure. In addition, our final computations will involve integration over spatial variables only. A spatial mesh therefore has the additional benefit of localising the integration regions, permitting the use of lower order quadrature rules.

The overall approximation is then of the form

$$\rho^\Gamma(X^s, \omega) \approx \sum_{\alpha=1}^{n} \sum_{l=0}^{N_s} \sum_{m=0}^{N_p} \rho_{\alpha,l,m} \hat{P}_{\alpha,l}(s) \tilde{P}_m(p_s), \tag{34}$$

where $N_s$ and $N_p$ are the orders of the position and momentum basis expansions, respectively, and $n$ is the number of elements in the mesh. Also

$$\hat{P}_{\alpha,l}(s) = \sqrt{2/A_\alpha} P_l(\tilde{s}_\alpha) \chi_\alpha(s),$$

where $\tilde{s}_\alpha$ parameterizes the $\alpha^{\text{th}}$ element and is scaled to take values in $(-1, 1)$, $\chi_\alpha$ is the characteristic function for element $\alpha$ and $A_\alpha$ is the length of element $\alpha$, $\alpha = 1, .., n$. An analogous approximation is also made for $\rho_0^\Gamma$, the initial density, and the values of $\rho_{\alpha,l,m}$ are to be determined by solving the resulting linear system. Note that the special case of piecewise constant boundary elements is reached by setting $N_s = 0$ above. The other limiting case, $N_p = 0$, corresponds to a 'diffusive' reflection approximation, also denoted as *Lambert's law* [5], which is equivalent to the assumption that incoming rays are scattered uniformly in all directions after reflecting from a boundary. Equation (30) then becomes equivalent to the integral equations discussed in [21, 30]. Note that the case $N_p = N_s = 0$ and $n = 1$ corresponds to an SEA approximation for a single subsystem ($n$ equals the number of substructures for multi-component systems ) [23].

The matrix approximation $B$ of the linear operator $\mathcal{B}$ (29) is given by taking the inner product $< \cdot, \cdot >$ in $L^2(\partial \mathbb{P})$:

$$
\begin{aligned}
&B_{m+1+N_p(l+N_s(\alpha-1)), b+1+N_p(a+N_s(\beta-1))} = \\
&\frac{(2m+1)(2l+1)}{4} \left\langle \mathcal{B}\left(\hat{P}_{\beta,a}(s')\tilde{P}_b(p'_s)\right), \hat{P}_{\alpha,l}(s)\tilde{P}_m(p_s) \right\rangle = \\
&\frac{(2m+1)(2l+1)}{4} \int_{\partial \mathbb{P}} \int_{\partial \mathbb{P}} w(Y^s,\omega)\delta(X^s-\gamma(Y^s))\hat{P}_{\beta,a}(s')\tilde{P}_b(p'_s)\hat{P}_{\alpha,l}(s)\tilde{P}_m(p_s) dY^s dX^s = \\
&\frac{(2m+1)(2l+1)}{4} \int_{\partial \mathbb{P}} w(Y^s,\omega)\hat{P}_{\beta,a}(s')\tilde{P}_b(p'_s)\hat{P}_{\alpha,l}(\gamma_s(Y^s))\tilde{P}_m(\gamma_p(Y^s)) dY^s.
\end{aligned}
$$
(35)

Here we write $\gamma = (\gamma_s, \gamma_p)$, to denote the splitting of the position and momentum parts of the boundary map. The boundary map can be very difficult to obtain in general, and hence we write the operator in terms of trajectories with fixed start and end points $s'$ and $s$ as follows

$$
\begin{aligned}
&B_{m+1+N_p(l+N_s(\alpha-1)), b+1+N_p(a+N_s(\beta-1))} = \\
&\frac{(2m+1)(2l+1)}{4} \int_\Gamma \int_\Gamma w(Y^s,\omega)\tilde{P}_m(p_s(s,s'))\hat{P}_{\alpha,l}(s)\tilde{P}_b(p'_s(s,s'))\hat{P}_{\beta,a}(s') \left| \frac{\partial p'_s}{\partial s} \right| ds' ds.
\end{aligned}
$$
(36)

The resulting boundary integral formulation containing a pair of integrals over boundary coordinates bears a resemblance to standard variational Galerkin boundary integral formulations, see for example [31].

## 5. Numerical results

In this section we consider a two-dimensional polygonal example whose boundary is meshed by subdividing each side into equidistant sections governed by a mesh parameter $\Delta x$. The number of elements on any given side is computed using the integer part of the side length divided by $\Delta x$. The Jacobian in Eq. (36) is written in the form

$$
\left| \frac{\partial p'_s}{\partial s} \right| = \frac{|p|(n \cdot (r-r'))(n' \cdot (r-r'))}{|r-r'|^3},
$$
(37)

where $n$ and $n'$ are the inward unit normal vectors to $\Gamma$ at $r$ and $r'$, respectively. In addition $r$ and $r'$ are the Cartesian coordinate vectors corresponding to the boundary parameterised coordinates $s$ and $s'$. In order to treat the corner singularities in (37), Gaussian quadrature is employed where end-points are not included as quadrature points. The convergence of the quadrature rules is still slow due to the peak in the integrand at corners. Telles' transformation techniques are employed to speed up the convergence [32].

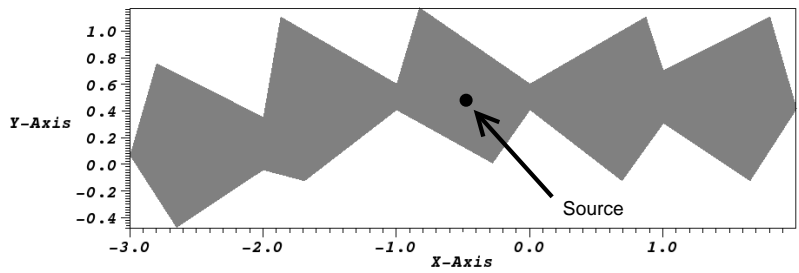

Figure 1: Five cavity polygonal example geometry

A five cavity system is considered as shown in Fig. 1 with Dirichlet boundary conditions on the outer boundary. The application of the method to multiple sub-domains is carried out using the same techniques as detailed in [24]. In general the configuration features irregular shaped, well separated pentagonal subsystems. In such a case one expects the wave energy to be fairly evenly distributed within subsystems and the energy levels in each subsystem to be distinct and well separated. The one feature of the configuration where such trends may not hold is due to the direct channel running to the right of the source point leading to a corridor type effect and energy being localised along this channel. Such dynamical features are likely to require higher order applications of our numerical algorithm to adequately resolve the behaviour.

Energy distributions have been studied as a function of the frequency with a hysteretic damping factor $\eta = 0.01$, where $\mu = \omega\eta/(2c)$ throughout $\Omega$. Here and in the remainder of this section the subsystems (or cavities) are numbered $1, 2, ..., 5$ from left to right. The wave speed is set to unity for simplicity. Note that the method allows $c$ (and thus $\mu$) to vary between subsystems [24], but we do not make use of this in the examples considered here. Therefore the subsystem interface reflection and transmission coefficients appearing in the weight term in (29) are simply 0 and 1, respectively.

Fig. 2 shows the energy distribution throughout the five subsystems compared with the results of a boundary element computation for the full wave problem. Explicitly we compute the square of the energy norm

$$\|G_i\|^2 := \int_{\Omega_i} |G(r, r_0; \omega)|^2 dr, \quad i = 1, 2, .., 5. \tag{38}$$

The lines show the approximate energy norms given by the Liouville equation model computed for $f = (\omega/(2\pi)) = 10, 13, 16, 19, ..., 70$ with $N_p = 10$, $N_s = 0$ and $n = 627$, and hence the spatial approximation has been improved via mesh refinement rather than using higher order Legendre polynomial basis expansions. This has the benefit that the quadrature costs are kept moderate and leads to considerable efficiency gains. The computations here took approximately 2.5 hours per frequency compared with typical times of approximately 53 hours for a full tenth order Chebyshev basis computation using the methods reported in [24]. Here the computations were performed using a single core of a desktop PC with a 2.83 Ghz dual core processor. The circles each show the mean of 22 boundary element method computations, where the average is taken over a small range of frequencies centered around 10, 20,..., 70. The plus signs show the maxima and minima of the boundary element computations. The plot shows a good agreement between the solution of the Liouville equation and the mean of the full wave problem computations. In addition, the Liouville solution always lies within the range of the data for the full wave problem.

Fig. 3 shows the convergence of our numerical algorithm in subsystem 5 as the precision of the approximation is increased by refining the mesh and simultaneously increasing the order of the Legendre basis approximation in direction space. The plot shows the relative error in the energy norm $\|G_5\|^2$ compared against the value $\|G_5^e\|^2$ computed with $N_p = 10$, $N_s = 0$ and $n = 627$. The relative error plotted is thus given by $(\|G_5^e\|^2 - \|G_5\|^2)/\|G_5^e\|^2$. The trend of the energy increasing as the precision of the numerical method is increased is shown by the error values all being positive and is expected since a low order implementation of our numerical scheme will not capture the direct energy channelled to the right of the source and into subsystem 5. Higher order basis functions in direction space are required to capture such directivity patterns in the wave field. In addition the plot shows the approximated solutions converging since the errors are getting closer to zero as higher order approximations are employed. For the case $N_p = 2$, $n = 70$ one sees the error increasing as the frequency and thus damping are increased. For high damping, the direct channel of energy (before reflection) from the source point will form a more significant part of the overall solution. For the case $N_p = 4$, $n = 70$, the higher order basis in direction space has been able to model the channelling effect more accurately and the error has become almost uniform across the frequency range.

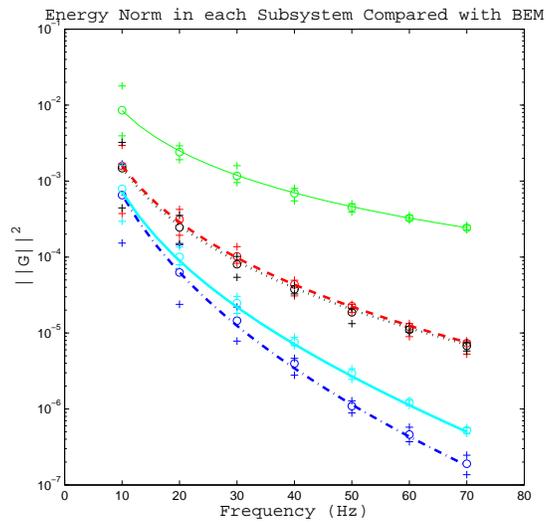

Figure 2: Energy distribution in five coupled acoustic cavities. Subsystem: 1 - blue (dash-dot), 2 - red (dashed), 3 - green (thin solid), 4 - black (dotted), 5 - cyan (thick solid) (colour online)

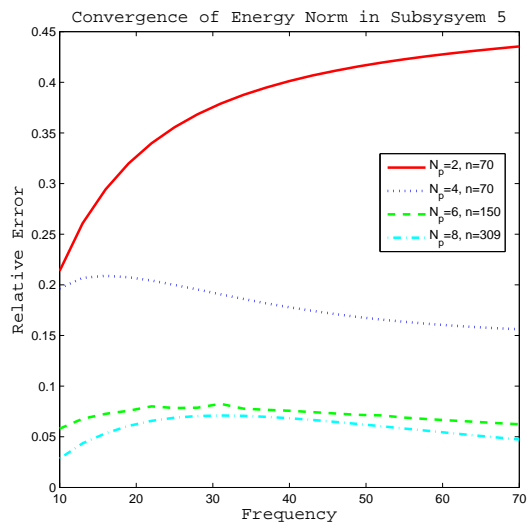

Figure 3: Evidence of convergence in subsystem 5 (colour online)

Fig. 4 shows the energy density $|G(r, r_0; \omega)|^2$ plotted for $r \in \Omega$ and $f = 15$. The upper subplot shows the energy density approximated using a boundary element method for the full wave problem. The central subplot shows the average of 22 full wave solutions at different frequencies $13 < f < 17$. One sees that the averaging reduces the influence of small scale fluctuations in the solution and produces a generally smoother looking plot, showing more clearly the influence of the source point. The lower subplot shows the energy density approximated using the Liouville equation model with $N_p = 10$, $N_s = 0$ and $n = 627$. Many similarities with the central subplot are apparent. For example, the direct energy channel to the right of the source and into subsystem 5, the location of the source and the dip in energy in the second subsystem close to the interface are features present in both plots. Also the energy channelled along the upper boundaries of subsystems 1 and 4 is a common feature. This demonstrates that high frequency wave models based on the Liouville equation are indeed a powerful tool to obtain averaged solutions. Such an average solution can be obtained through averaging over an ensemble of similar structures or over a small frequency range. Applications are abundant in high frequency vibro-acoustics, where small changes in the domain due to, for example, manufacturing tolerances lead to large variations in the response and thus a simulation of any one particular problem will only be of statistical significance. Indeed, replacing highly oscillatory wave problems with smooth ray dynamical problems is often the only practical option in the high frequency regime.

## 6. Conclusion

The suitability of the Liouville equation as a model for high frequency wave problems has been discussed. A boundary integral approach has been proposed for its solution using the Frobenius-Perron operator. A numerical solution method has been implemented based on Legendre polynomial approximations in phase space together with a spatial boundary mesh. The proposed method has been verified by comparison with a BEM code for the full wave problem and numerical evidence of convergence has been shown.

## ACKNOWLEDGEMENT

Support from the EPSRC (grant EP/F069391/1) and the EU (FP7IAPP grant MIDEA) is gratefully acknowledged. The authors also wish to thank InuTech Gmbh, Nürnberg for Diffpack guidance and licences, and Dmitrii Maksimov and Niels Søndergaard for stimulating discussions.

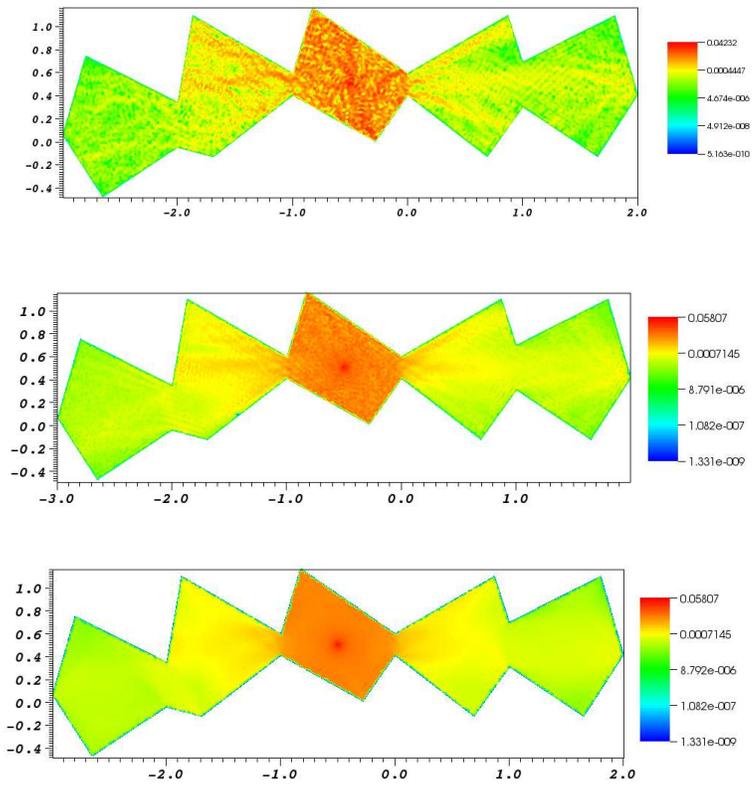

Figure 4: Energy density compared against the full wave solution: upper subplot shows the full wave solution, central subplot shows the full wave solution averaged over a small frequency band, lower subplot shows the Liouville equation model (colour online)